\documentclass[slac_one]{revtex4}
\usepackage{graphicx}
\usepackage{fancyhdr}
\begin{document}

\newcommand{\unit}[1]{\,{\mathrm{#1}}}
\newcommand{\as}{\overline{\alpha}_s}
\newcommand{\epem}   {\ensuremath{\mathrm{e^+e^-}}}
\newcommand{\roots} {\sqrt{s}}
\newcommand{\ecm} {\ensuremath{E_{\rm cm}}}
\newcommand{\mz}     {\ensuremath{M_{\rm Z}}}
\newcommand{\mw}     {\ensuremath{M_{\rm W}}}
\newcommand{\bt}     {\ensuremath{B_T}}
\newcommand{\bn}     {\ensuremath{B_N}}
\newcommand{\bw}     {\ensuremath{B_W}}
\newcommand{\mh}     {\ensuremath{M_h^2/s}}
\newcommand{\mhd}    {\ensuremath{(M_h^2-M_l^2)/s}}
\newcommand{\thr}    {\ensuremath{1-T}}
\newcommand{\ptin}    {\ensuremath{p_{\perp}^{\rm in}}}
\newcommand{\ptout}    {\ensuremath{p_{\perp}^{\rm out}}}
\newcommand{\nch}    {\ensuremath{\langle N_\mathrm{ch}\rangle}}
\newcommand{\tmaj}   {\ensuremath{T_{\mathrm{major}}}}
\newcommand{\tmin}   {\ensuremath{T_{\mathrm{minor}}}}
\newcommand{\oaa}    {\ensuremath{\mathcal{O}(\alpha_s^2)}}
\newcommand{\alphab} {{\overline \alpha_s}}
\newcommand{\xp}     {\ensuremath{x_p}}
\newcommand{\xmu}     {\ensuremath{x_\mu}}
\newcommand{\mui}     {\ensuremath{\mu_{\rm I}}}
\newcommand{\ksip}   {\ensuremath{\xi}}
\newcommand{\klphd}   {\ensuremath{K_{\rm LPHD}}}
\newcommand{\xistar} {\ensuremath{\xi^*}}
\def\half{\mbox{\small $\frac{1}{2}$}}
\newcommand{\la}{\langle}
\newcommand{\ra}{\rangle}
\newcommand{\myn}    {\ensuremath{\langle y^n\rangle}}
\newcommand{\vecma}  {\ensuremath{\vec{n}_{\mathrm{Ma}}}}
\newcommand{\vecmi}  {\ensuremath{\vec{n}_{\mathrm{Mi}}}}
\newcommand{\vecsma}  {\ensuremath{\vec{n}_{\mathrm{sMa}}}}
\newcommand{\vecsmi}  {\ensuremath{\vec{n}_{\mathrm{sMi}}}}
\newcommand{\chidof}     {\ensuremath{\chi^2/N_{\rm DOF}}}
\newcommand{\evis}     {\ensuremath{E_{\rm vis}}}
\newcommand{\ycut}     {\ensuremath{y_{\rm cut}}}
\newcommand{\ymax}     {\ensuremath{y_{\rm max}}}
\newcommand{\stot}     {\ensuremath{\sigma_{\rm tot}}}
\def\Covrln{{\overline C}}
\def\Govrln{{\overline G}}
\def\Lovrln{{\overline L}}
\def\Lovrlntilde{{\widetilde{\overline L}}}
\def\Lbar{{\bar L}}
\def\as{\alpha_{s}}
\def\gae{{\gamma_{\textsc{e}}}}
\def\asb{{\bar \alpha}_{{\textsc{s}}}}

\title{Multijets in $\epem$ annihilation}

%

\author{H.~Stenzel}
\affiliation{University of Giessen, II. Physikalisches Institut, Heinrich-Buff Ring 16, D-35392 Giessen, Germany}

\begin{abstract}
A review of multijet rates and event-shape variables sensitive to 
multijet configuration with applications to measurements of $\alpha_s$, 
QCD color factors and tests of power corrections is presented. 

\end{abstract}

\maketitle

\thispagestyle{fancy}


\section{Introduction}
A wealth of results on QCD in $\epem$ annihilation emerged from the 
LEP experiments. The main emphasis of these studies was on high-statistics 3-jet 
configurations. However, in the second part of LEP running and with the advent 
of NLO calculations for $\epem \rightarrow 4 $jets, a new class of studied 
appeared yielding high precision results using multijet events. 

\begin{figure}[h]
\includegraphics[width=100mm]{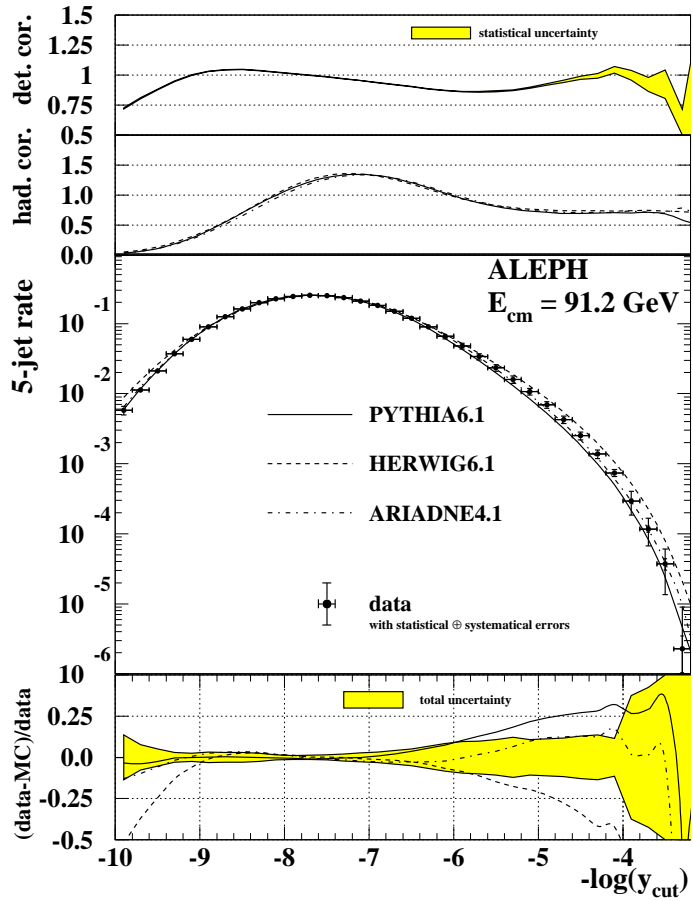}
\caption{The 5-jet rate measured at LEP1 using the Durham jet algorithm.}
\label{fig:aleph5j}
\end{figure}

The most basic observable characterising  multijet configurations are 
jet rates. As an example the 5-jet rate measured at LEP1 by ALEPH \cite{aleph_qcd} 
is shown in Fig.~\ref{fig:aleph5j}.   
The presented data are corrected for acceptance and detector resolution effects to 
the hadron level. Hadronisation corrections, relevant for comparisons with NLO 
calculations, are not applied at this stage but shown in Fig.~\ref{fig:aleph5j} to 
indicate their size. The experimental systematic uncertainties are between one and 
two $\%$ in the central part and increasing toward the phase space edges. The measurements 
are compared to various leading-log generators, which generate higher jet multiplicity 
only in the cascade of parton showers. Again in the peak region a reasonable 
agreement is observed.

\begin{figure}[h]
\includegraphics[width=100mm]{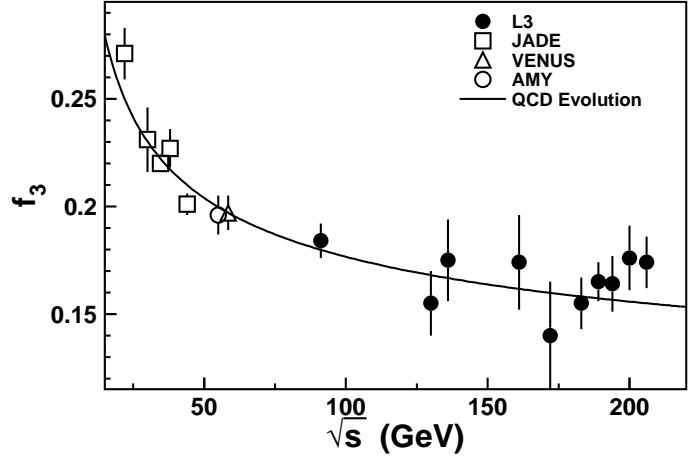}
\caption{The evolution of the 3-jet rate with $\ecm$}
\label{fig:l3j}
\end{figure}

Jet rates exhibit a clear dependence on the value of $\alpha_s$, since 
at leading order $R_{n+2} \propto \alpha_s^n$. The evolution of jet rates 
at a fixed value of $\ycut$  
with centre-of-mass energy hence reveals the running of $\alpha_s$, 
as demonstrated by a measurement of L3 \cite{l3_qcd} shown in 
Fig.~\ref{fig:l3j}.  

\section{4-jet observables}
Naturally the class of observables receiving most of attention beyond 
3-jets are the 4-jet observables. 4-jet production receives contributions 
from double-bremsstrahlung and gluon splitting diagrams, the latter into 
gluonic and quarkonic final states, each of them contributing with 
a different color factor (ratio). This color structure can be measured 
by combining different angular observables constructed in the 
4-jet system in order to extract the color factor ratios 
$C_A/C_F$ and $T_R/C_F$ with nominal QCD SU(3) values $C_A$=3, $C_F=4/3$ and 
$T_R=1/2$.  On top of this residual dependence on the color factors 
the perturbative predictions for 
differential distributions receive an overall normalisation factor in 
powers of $\alpha_s$, yielding at NLO a prediction of the following form:
\begin{equation}\label{eq:O4}
\frac{1}{\sigma}\frac{d\sigma}{dO_4} = \eta(\mu)^2 B(O_4) + 
\eta(O_4)^3\left[B(O_4)\beta_0\ln x_\mu^2 + C(O_4)\right] + {\cal O}(\eta^4)\; ; \; \; 
\eta(\mu)=\frac{\alpha_s(\mu)\cdot C_F}{2\pi}\; ,
\end{equation}
for a generic 4-jet observable $O_4$. Therefore, the analyses determine 
usually the color factors and $\alpha_s$ simultaneously, using fits to 
both angular variables with a particular sensitivity to the color factor ratios 
and event-shape- / jetrate-type observables with better sensitivity to $\alpha_s$. 
The distributions of representative observables of each class are shown for 
the angular correlations in Fig.~\ref{fig:a34} and for the 4-jet rate in Fig.~\ref{fig:ar4}.    
\newlength{\wi}   \wi 0.48\textwidth
\newlength{\fwi}  \fwi 0.9\wi
\newlength{\hwi}  \fwi 0.9\wi
\begin{figure}[h]
\begin{minipage}[b]{\wi}
\centering\includegraphics[width=\fwi]{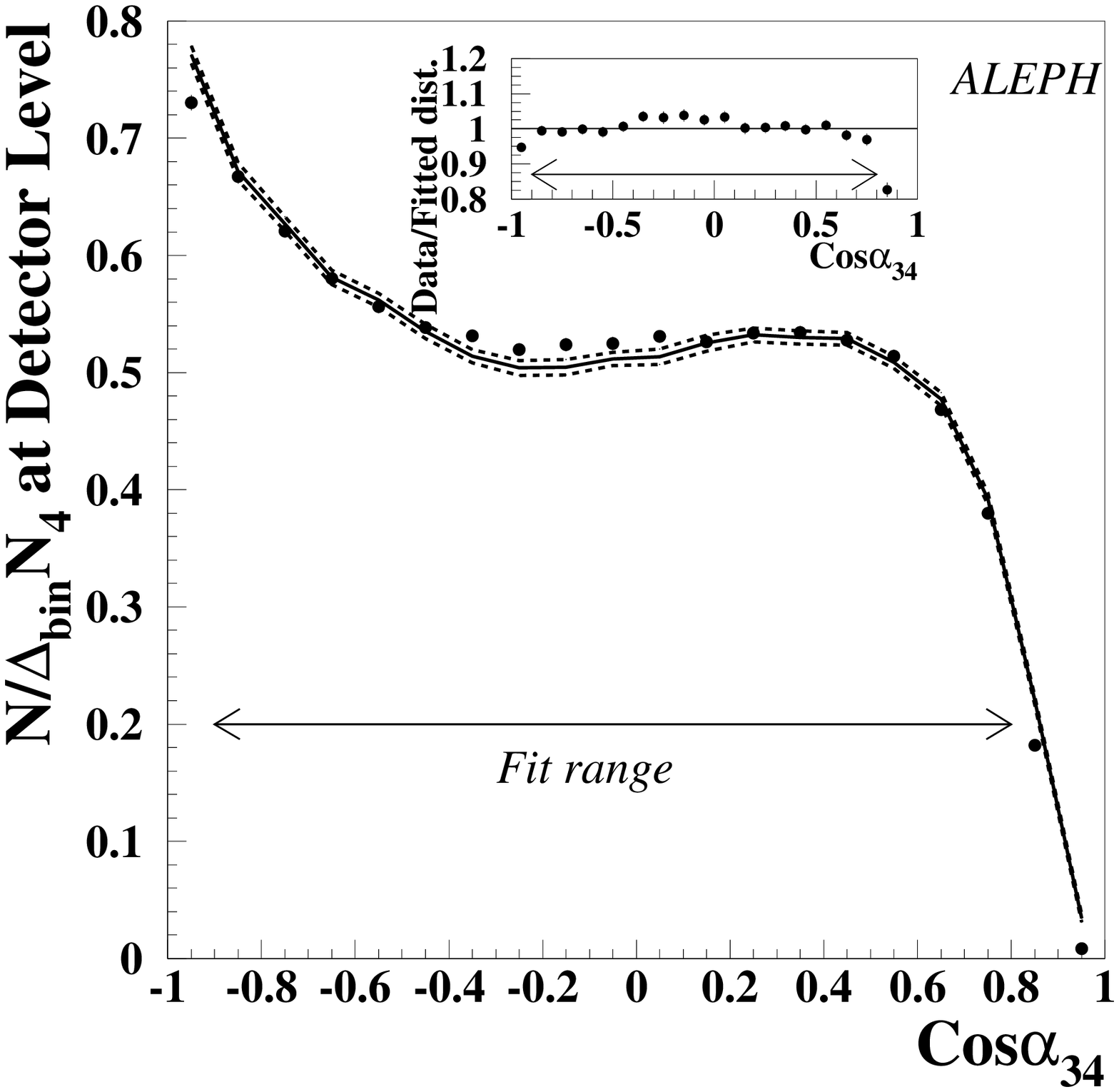}
\caption{Distribution of the angle between the two lowest energetic jets $\alpha_{34}$ 
compared to the NLO prediction.}
\label{fig:a34}
\end{minipage}\hfill
\fwi 0.91\wi
\begin{minipage}[b]{\wi}
\centering\includegraphics[width=\fwi]{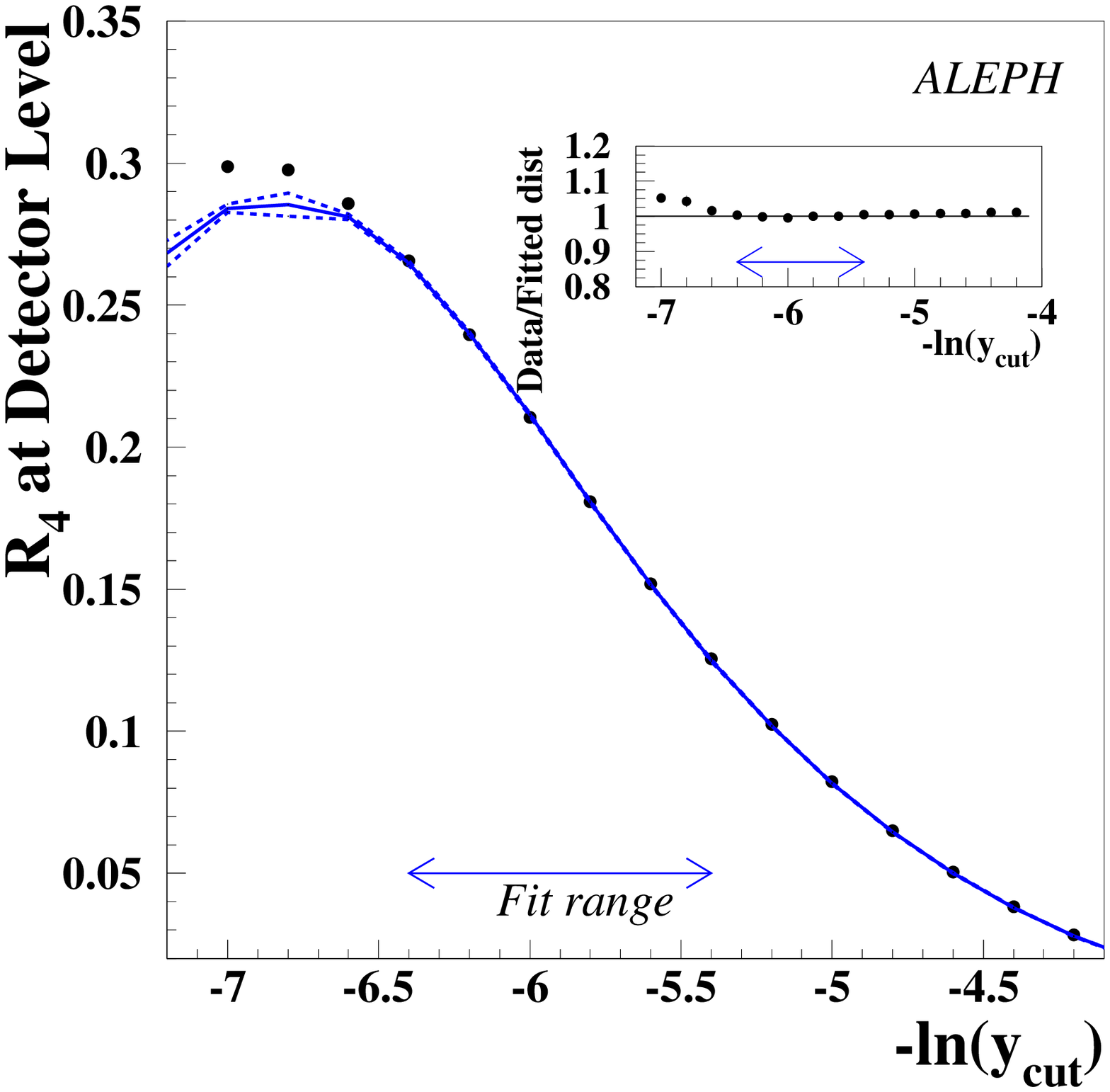}
\caption{The 4-jet rate measured by ALEPH 
compared to the NLO+NLLA prediction.}
\label{fig:ar4}
\end{minipage}
\end{figure}

\begin{figure}[h]
\includegraphics[width=80mm]{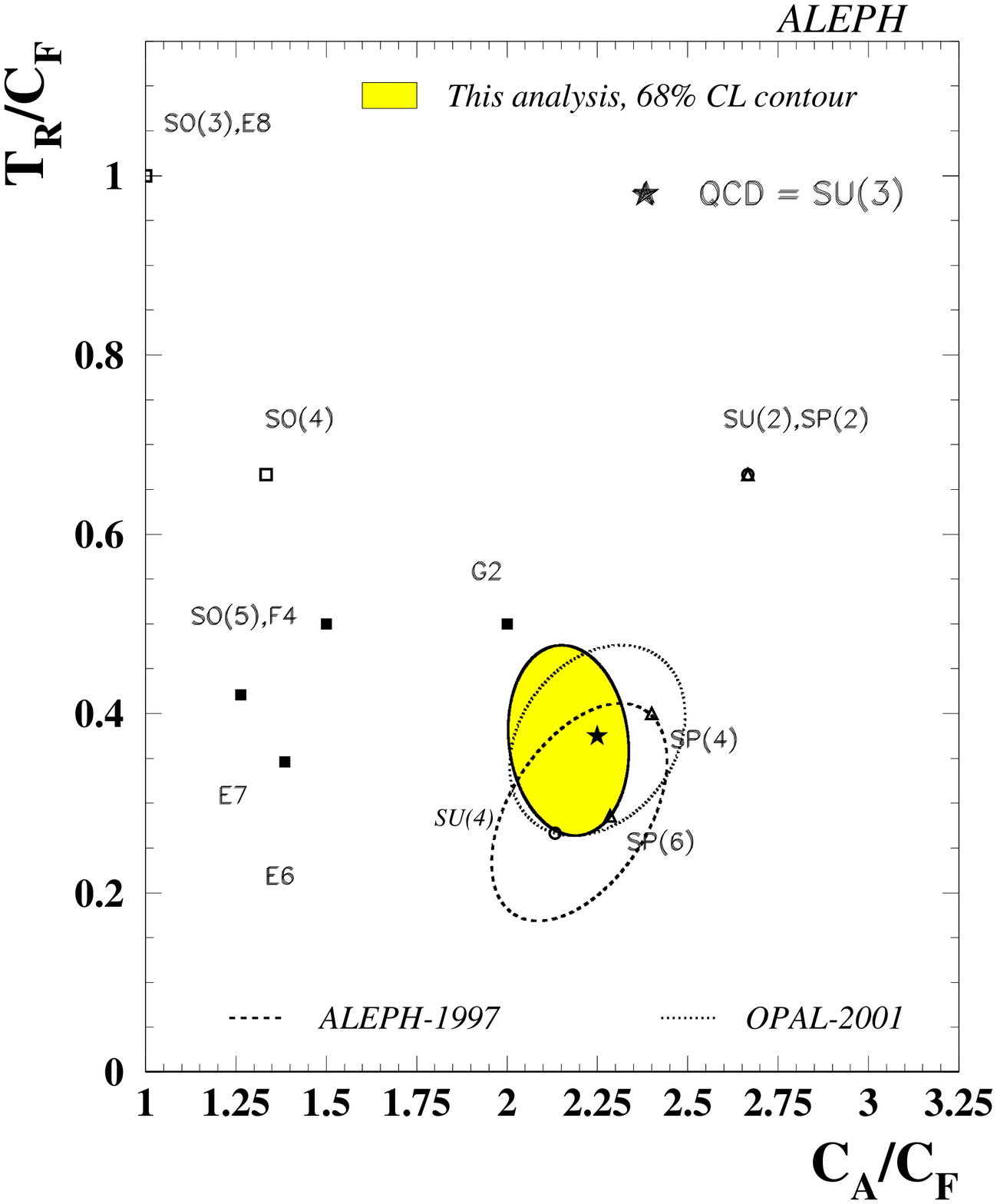}
\caption{CL contours in the plane of $T_R/C_F$ versus $C_A/C_F$.}
\label{fig:cacf}
\end{figure}

\begin{figure}[h]
\begin{minipage}[b]{\wi}
\centering\includegraphics[width=\fwi]{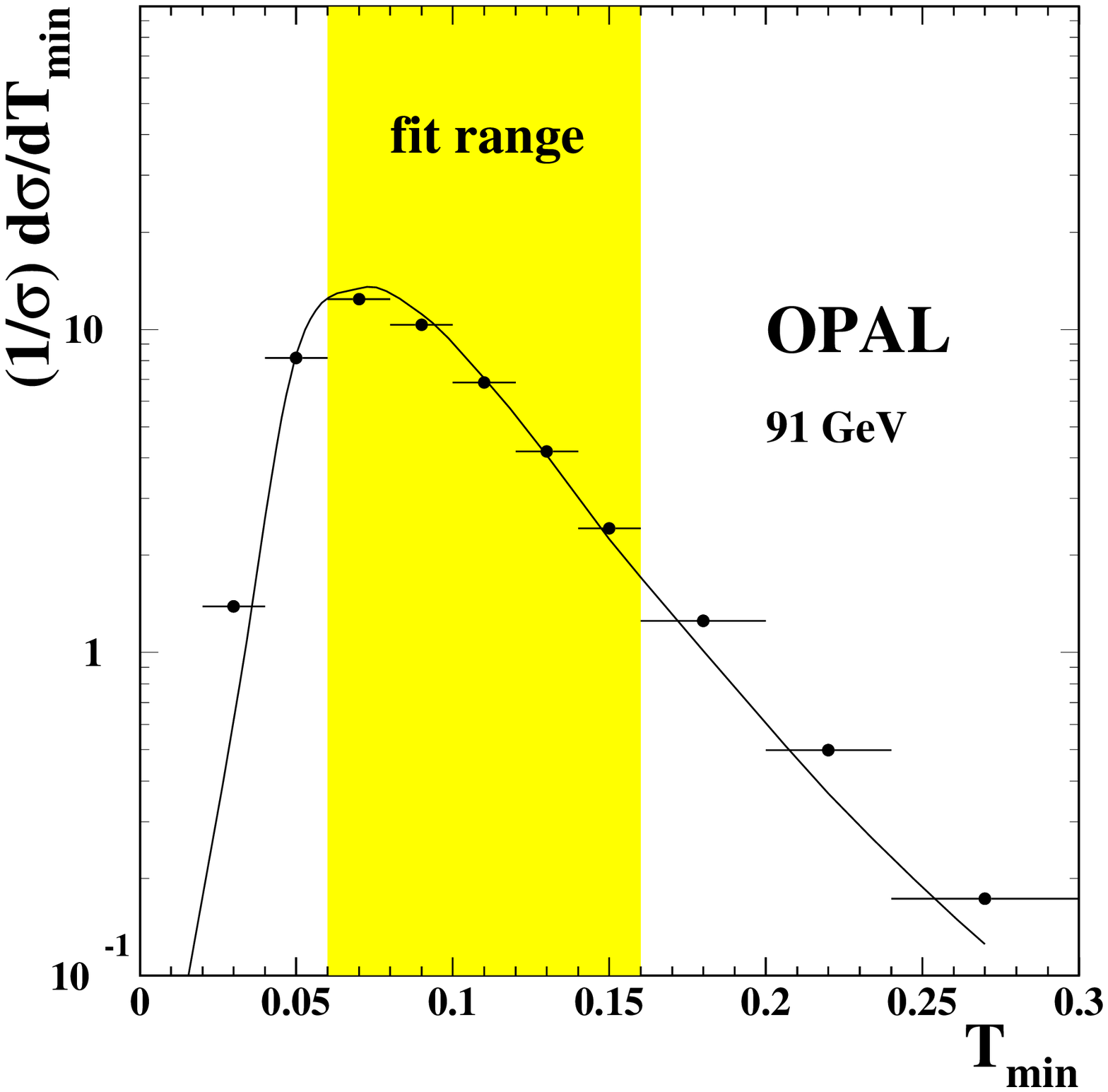}
\caption{Distribution of $T_{min}$ measured by OPAL  
compared to the NLO prediction.}
\label{fig:otmin}
\end{minipage}\hfill
\fwi 0.91\wi
\begin{minipage}[b]{\wi}
\centering\includegraphics[width=\fwi]{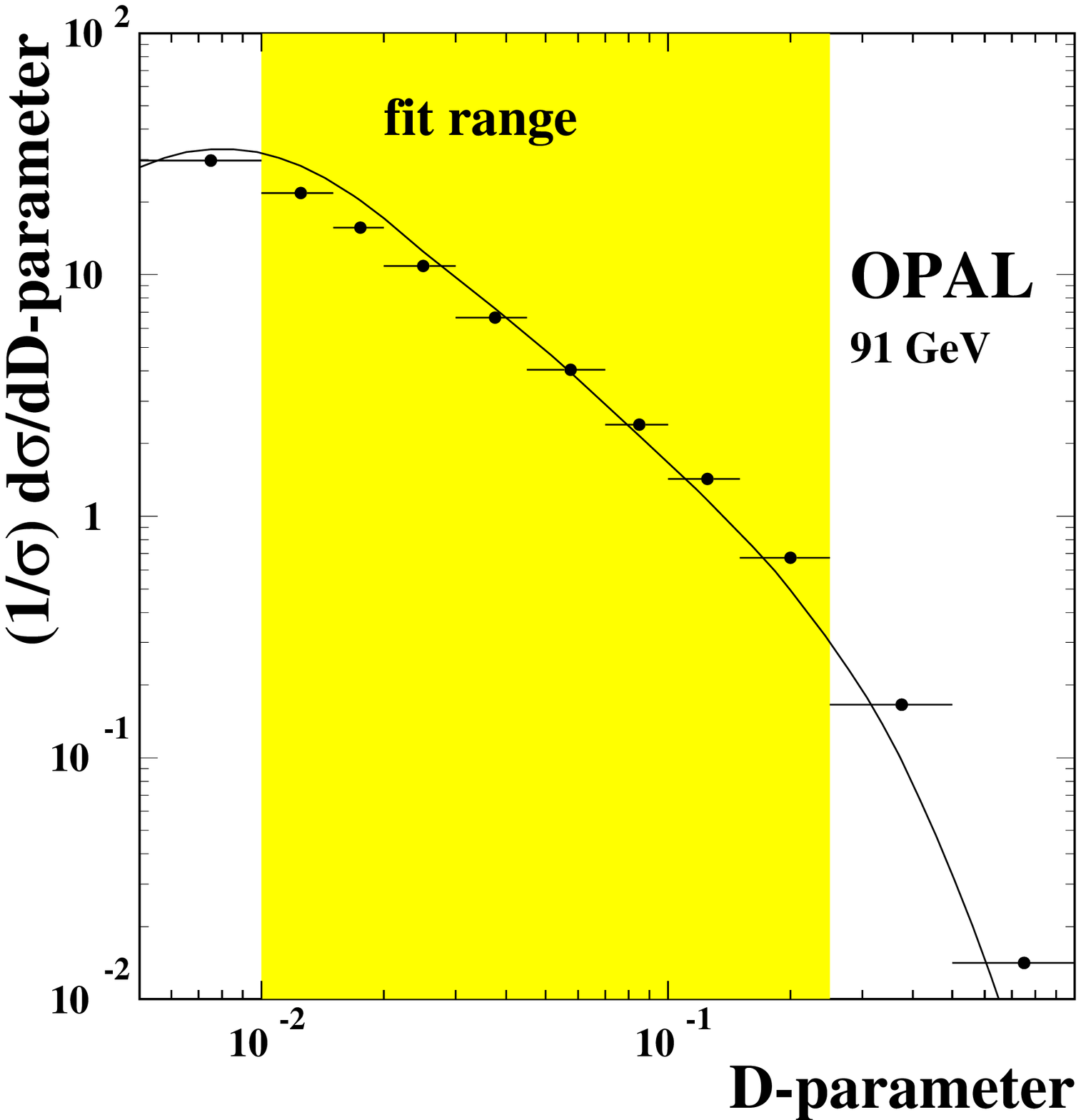}
\caption{Distribution of $T_{min}$ measured by OPAL  
compared to the NLO prediction. }
\label{fig:odpar}
\end{minipage}
\end{figure}

Simultaneous analyses have been carried out recently by ALEPH \cite{aleph_colfac} 
and OPAL \cite{opal_colfac} yielding results summarised in Table~\ref{tab:colfac}.  
\begin{table}[h]
\begin{center}
\begin{tabular}{|l|l|l|l|}
\hline &  \bf{$\alpha_s(M_Z)$} & \bf{$C_A$} & \bf{$C_F$} \\ \hline
ALEPH & $0.119 \pm 0.006 \pm 0.026 $ & $2.93 \pm 0.14 \pm 0.58 $ & $ 1.35 \pm 0.07 \pm 0.26$ \\ 
OPAL & $0.120 \pm 0.011 \pm 0.020 $ & $3.02 \pm 0.25 \pm 0.49 $ & $ 1.34 \pm 0.13 \pm 0.22$ \\
QCD &  & $ 3 $ & $ 4/3$ \\ \hline
\end{tabular}
\caption{Results of NLO analyses of 4-jet observables, the first error is statistical 
and the second systematic.}
\label{tab:colfac}
\end{center}
\end{table}
The results for the color factors are in excellent agreement with the expectation 
of SU(3) for QCD, as shown in Fig.~\ref{fig:cacf}.  
The precision in $\alpha_s$ obtained from simultaneous fits is rather limited though 
consistent with other determinations. Given the agreement of the color factor measurements 
with the QCD expectation, one can also assume the latter and extract $\alpha_s$ alone 
from 4-jet observables. It turns out that the systematic uncertainty of such 
a measurement is better than from 3-jet observables, given the quadratic dependence 
on $\alpha_s$ at lowest order 
\begin{equation}
\frac{\Delta \alpha_s}{\alpha_s} \approx \frac{1}{2}\frac{\Delta O_4}{O_4} \; .
\end{equation} 
OPAL recently presented an analysis \cite{opal_4jets} of various 4-jet observables, including 
the 4-jet event-shape variables thrust minor (Fig.~\ref{fig:otmin}) and D-parameter (Fig.~\ref{fig:odpar}) along these lines, using data 
at LEP1 and LEP2. 
On top of the NLO calculation, for one observable, the 4-jet rate in the Durham scheme, 
also an all-orders resummation is available. As observed already in the 3-jet 
case, this observable yields the best precision in terms of systematic uncertainty. 

\begin{figure}[h]
\includegraphics[width=100mm]{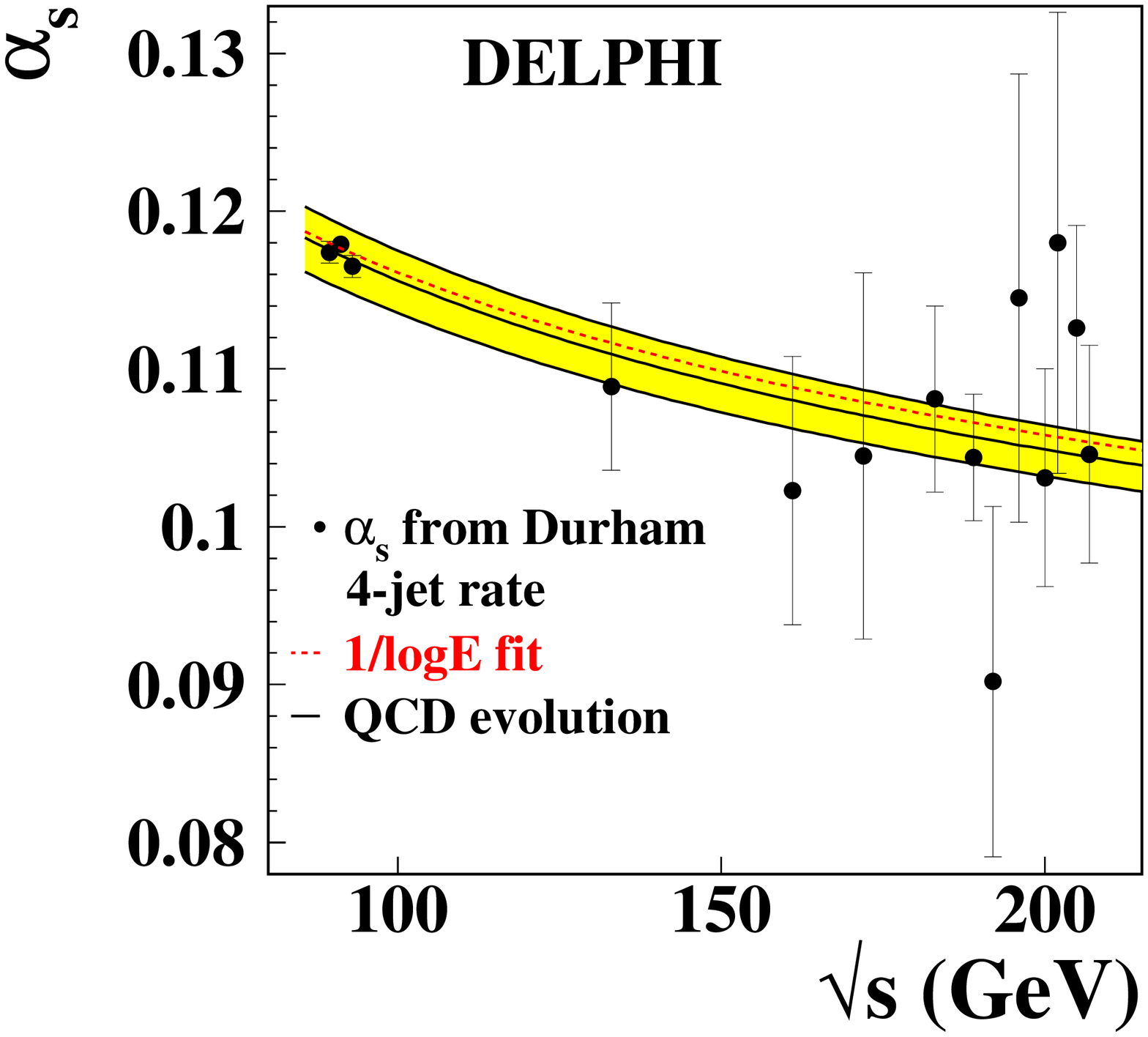}
\caption{A test of the energy evolution of $\alpha_s$ determined 
from the 4-jet rate performed by DELPHI.}
\label{fig:d4jet}
\end{figure}

A similar work has also been published by DELPHI \cite{delphi_4jets} using 
pure NLO with optimised scales. DELPHI turned the measurement of $\alpha_s$ 
from the 4-jet rate into a test of the logarithmic slope $d\alpha_s^{-1}/d\log E$ of the 
energy evolution governing the running of $\alpha_s$, shown in Fig.~\ref{fig:d4jet}.  
 
All of these analyses agree in their competitive precision compared to 
other determinations of $\alpha_s$ mainly resulting from a reduced 
perturbative uncertainty. A summary of the results on $\alpha_s$ 
from the 4-jet rate obtained by the LEP collaborations is given in Table~\ref{tab:as4j}. 
\begin{table}[h]
\begin{center}
\begin{tabular}{|l|l|l|l|l|}
\hline &  \bf{$\alpha_s(M_Z) \pm \Delta_{stat}$} & \bf{$\Delta_{sys}$} & \bf{data set} & \bf{theory} \\ \hline
OPAL  & $0.1182 \pm 0.0003 $ & $0.0026$ & 91-209 & NLO+NNLA \\
ALEPH & $0.1170 \pm 0.0001 $ & $0.0022$ & 91 & NLO+NNLA \\
DELPHI & $0.1175 \pm 0.0005 $ & $0.0030$ & LEP1 & NLO+$x_\mu^{opt}$ \\ \hline
DELPHI & \multicolumn{2}{c}{Logarithmic slope = $1.14\pm 0.36$} & 89-209 & QCD: $1.27\pm 0.10$ \\ \hline
\end{tabular}
\caption{Summary of results on $\alpha_s$ from the 4-jet rate.}
\label{tab:as4j}
\end{center}
\end{table}

\section{Power corrections for multijets}
\begin{figure}[h]
\includegraphics[width=95mm]{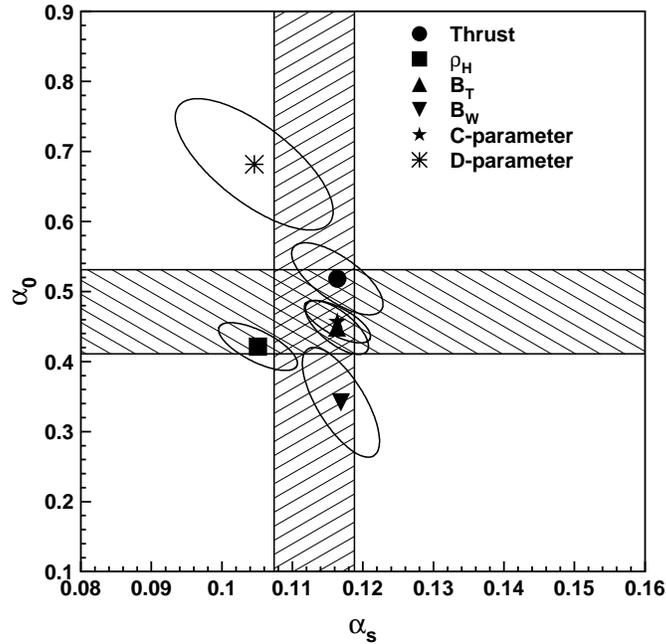}
\caption{Confidence level contours in the plane of $\alpha_0$ vs. $\alpha_s$ 
for power correction analyses to event-shape mean values by L3.}
\label{fig:l3asa0}
\end{figure}

Most of the effort in understanding non-perturbative effects in $\epem$ annihilations 
went into 3-jet observables, as highlighted in various contributions to this 
conference. The most common case also applied by the experiments is the power 
law correction, where the non-perturbative correction is absorbed in to 
a term scaling with powers of $1/Q$ and parameterised by the moment of 
an effective coupling $\alpha_0$ evaluated at a scale $\mu_I$ usually set to 2 GeV.  
With the availability of NLO calculations for 4-jet observables 
and the calculation of the corresponding power correction terms, 
the scope of studies has been extended by  
an analysis of the D-parameter carried out by L3~\cite{l3_qcd}. In this 
case the full prediction including perturbative and power correction terms 
is given by
\begin{eqnarray}
<D> = <D_{pert}> + <D_{pow}>\; , \\ \nonumber
<D_{pert}> = B_D\left(\frac{\alpha_s}{2\pi}\right)^2 + D_D\left(\frac{\alpha_s}{2\pi}\right)^3 \; , \\ \nonumber
<D_{pow}> = 195 \frac{\alpha_s}{2\pi} {\cal P}(1/Q) \; .
\end{eqnarray} 
L3 performed a study of the mean values of event-shape variables measured over a large 
range of centre-of-mass energies and determined in simultaneous fits $\alpha_s$ and 
$\alpha_0$. The results of this analysis are summarised in Table~\ref{tab:l3pc}.
\begin{table}[h]
\begin{center}
\begin{tabular}{|l|l|l|}
\hline L3  &  \bf{$\alpha_s(M_Z)$} & $\alpha_0$(2 GeV) \\ \hline
D-parameter & $0.1046 \pm 0.0078 \pm 0.0096 $ & $0.682 \pm 0.094 \pm 0.018 $ \\ 
all combined & $0.1126 \pm 0.0045 \pm 0.0039 $ & $0.478 \pm 0.054 \pm 0.024 $ \\ \hline
\end{tabular}
\caption{L3 analysis of power corrections from event-shape means, the first error is statistical 
and the second systematic.}
\label{tab:l3pc}
\end{center}
\end{table}
The results from the D-parameter turn out to be only marginally 
consistent with overall combination (dominated by 3-jet event-shapes). The value 
of $\alpha_s$ is lower and the one of $\alpha_0$ much higher. This can also bee seen 
in the confidence level contour plot in Fig.~\ref{fig:l3asa0}. The systematic uncertainties 
for the measurement using the D-parameter are also significantly larger. This seems to indicate 
that, although a reasonable qualitative description is obtained, additional terms 
are needed to achieve a good quantitative description. 

\section{Conclusions}
Multijet observables, mostly based on 4-jet configurations, 
have given access to precision measurements of $\alpha_s$ and the QCD 
color factors. A wealth of experimental data on 4-jet event-shape variables 
is available from LEP and awaiting higher order QCD calculations 
to be analysed. Currently, only the D-parameter has been analysed for 
power corrections yielding mildly discrepant results. This is to be confirmed 
by further analyses of other 4-jet event shapes. Beyond 4-jets, 
jet rates have been measured for up to 6 jets and for certain 
multijet-configurations dedicated observables like the 
jet-resolution parameter $y_{ij}$ are available for 5- and 6-jets. 
The potential of these high jet multiplicity events is still to 
be explored.  
\begin{acknowledgments}
I wish to thank the organisers of this workshop G.~Salam, M.~Dasgupta and 
Y.~Dokshitzer for their support and the opportunity to present 
results from the LEP collaborations at this occasion.
\end{acknowledgments}


\end{document}